\documentclass{optica-article}

\journal{opticajournal} 

\articletype{Research Article}

\usepackage{lineno}
\usepackage{siunitx}
\usepackage{pifont}

\usepackage{amsmath}

\begin{document}

\title{A deep learning-based noise correction method for light-field fluorescence microscopy}

\author{Bohan Qu,\authormark{1,$\dag$} Zhouyu Jin,\authormark{2,$\dag$} You Zhou,\authormark{2,*} Bo Xiong,\authormark{3,*}and Xun Cao\authormark{1}}

\address{\authormark{1}Nanjing University, School of Electronic Science and Engineering, Nanjing, 210023, China\\
\authormark{2}Nanjing University, Medical School, Nanjing, 210093, China\\
\authormark{3}Peking University, National Key Laboratory of Multimedia Computing, Beijing, 100871, China\\
\authormark{$\dag$}These authors contributed equally.
}

\email{\authormark{*}You Zhou: zhouyou@nju.edu.cn, Bo Xiong: xiongbo@pku.edu.cn} 


\begin{abstract*} 
Light-field microscopy (LFM) enables rapid volumetric imaging through single-frame acquisition and fast 3D reconstruction algorithms. The high speed and low phototoxicity of LFM make it highly suitable for real-time 3D fluorescence imaging, such as studies of neural activity monitoring and blood flow analysis. However, in vivo fluorescence imaging scenarios, the light intensity needs to be reduced as much as possible to achieve longer-term observations. The resulting low signal-to-noise ratio (SNR) caused by reduced light intensity significantly degrades the quality of 3D reconstruction in LFM. Existing deep learning-based methods struggle to incorporate the structured intensity distribution and noise characteristics inherent to LFM data, often leading to artifacts and uneven energy distributions. To address these challenges, we propose the denoise-weighted view-channel-depth (DNW-VCD) network, integrating a two-step noise model and energy weight matrix into an LFM reconstruction framework. Additionally, we developed an attenuator-induced imaging system for dual-SNR image acquisition to validate DNW-VCD’s performance. Experimental results our method achieves artifact-reduced, real-time 3D imaging with isotropic resolution and lower phototoxicity, as verified through imaging of fluorescent beads, algae, and zebrafish heart. 

\end{abstract*}

\section{Introduction}
 Unveiling the dynamic movement and spatiotemporal distribution of cells is crucial for advancing current biological and medical research.\cite{RN1,RN2} Many biological processes occur in three-dimensional (3D) space on a millisecond timescale,\cite{RN3,RN4,RN5} posing significant challenges for rapid 3D microscopy. While 3D microscopy techniques such as confocal microscopy,\cite{RN6,RN7} light-sheet microscopy,\cite{RN8,RN9,RN10} and two-photon microscopy\cite{RN11,RN12} have proven effective, light-field microscopy (LFM)\cite{RN13,RN14,RN15,RN16} has recently emerged as a more convenient and rapid volumetric imaging method. LFM encodes 3D volumetric information through single-frame acquisition and reconstructs the 3D volume using specialized algorithms.\cite{RN14,RN17,RN18,RN19,RN20} With advantages such as low photobleaching and phototoxicity, LFM is well-suited for long-term, real-time 3D fluorescence observations of live biological samples, making it particularly valuable for high-speed imaging of neural activity,\cite{RN15,RN21,RN22} blood flow analysis,\cite{RN17,RN23} and other biological phenomena.\cite{RN17,RN24} 

With the advent of deep learning in optical microscopy, the focus of LFM reconstruction has shifted from traditional digital refocusing\cite{RN13} and deconvolution methods\cite{RN14,RN15,RN25} to learning-based approaches.\cite{RN17,RN20,RN26,RN27} These deep learning techniques\cite{RN17,RN20,RN26,RN27} have demonstrated superior performance in artifact removal, resolution enhancement, and optical aberration correction, reconstructing 3D volume by extracting multi-view images. However, the volumetric reconstruction performance of LFM is highly influenced by the signal-to-noise ratio (SNR) of the captured images. Therefore, LFM performance often falls short when requiring reduced light intensity and shorter exposure times, such as high-speed or light-sensitive in vivo imaging.\cite{RN28} Without explicit noise modeling, current methods often yield uneven energy distribution in reconstruction, particularly when applied to light-field images with extremely low SNR. Moreover, the structured grid-like intensity distribution of light-field images, formed by the MLA, distinguishing it from most other imaging modalities. Consequently, directly applying existing deep learning-based denoising strategies\cite{RN29,RN30,RN31} or adaptive noise model-based sparse filtering techniques\cite{RN32,RN33} to light-field data may compromise structural integrity and degrade reconstruction quality. Thus, existing deep learning-based LFM reconstruction methods fail to effectively incorporate the noise model and structural distribution characteristics of light-field data into their frameworks.

In this work, we propose the denoise-weighted view-channel-depth (DNW-VCD) network for LFM reconstruction. Our approach integrates a two-step noise model and an energy weight matrix for multi-view images into the existing VCD network framework\cite{RN17}. Specifically, beyond addressing camera pattern noise,\cite{RN32} we introduce a noise estimation strategy based on the characteristics of the microlens array (MLA) mask and an energy weight matrix derived from the intensity distribution patterns of multi-view sub-images in light-field data. By precisely modeling and approximating real-world noise distributions in LFM, our method significantly improves the network's denoising performance. Additionally, we develop an attenuator-induced imaging hardware for LFM, enabling flexible control over illimitation intensity. This hardware setup facilitates the simultaneous capture of dual-SNR light-field images, which we leverage to validate the accuracy of our method in dynamic imaging scenarios. Our results demonstrate that DNW-VCD allows for isotropic resolution, reduced artifacts, and real-time 3D reconstruction, even in extremely low SNR conditions. Our method markedly enhances LFM’s capability for low-light 3D imaging of biological samples, offering improved biocompatibility by minimizing photobleaching and phototoxicity. To validate the effectiveness and practicality of DNW-VCD, we conduct extensive experiments, including simulations with fluorescent beads under varying noise levels, low-light 3D imaging of biological slice samples, real fluorescent beads in water, and live zebrafish hearts. We also image free-moving algae under simultaneous high- and low-SNR conditions using a 20×/0.5 NA water-immersion objective, preliminarily investigating algal activity under different light intensities. These results confirm the accuracy and biocompatibility of our method, enabling volumetric imaging at a scale of 250 × 250 × 101 $\si{\micro\meter}^{3}$ with a temporal resolution of 70 fps. 

\section{Methods}

\subsection{\label{sec:level2}DNW-VCD principle}

In the DNW-VCD framework, the system imaging noise is modeled as the superposition of the camera pattern noise and Gaussian noise.\cite{RN32} This two-step noise model, along with a weight matrix of multi-view images, is integrated into the LFM reconstruction network to enhance denoising performance. To validate the accuracy of DNW-VCD in dynamic scenarios, an attenuator-controlled imaging system is developed to capture light-field images at adjustable high-to-low SNRs. The following sections provide an overview of the optical setup and the DNW-VCD principle, with additional details available in the Supplementary.
\subsection{Optical set-up of LFM and attenuator-induced imaging modality}
\begin{figure}[htbp]
\includegraphics[scale=0.82]{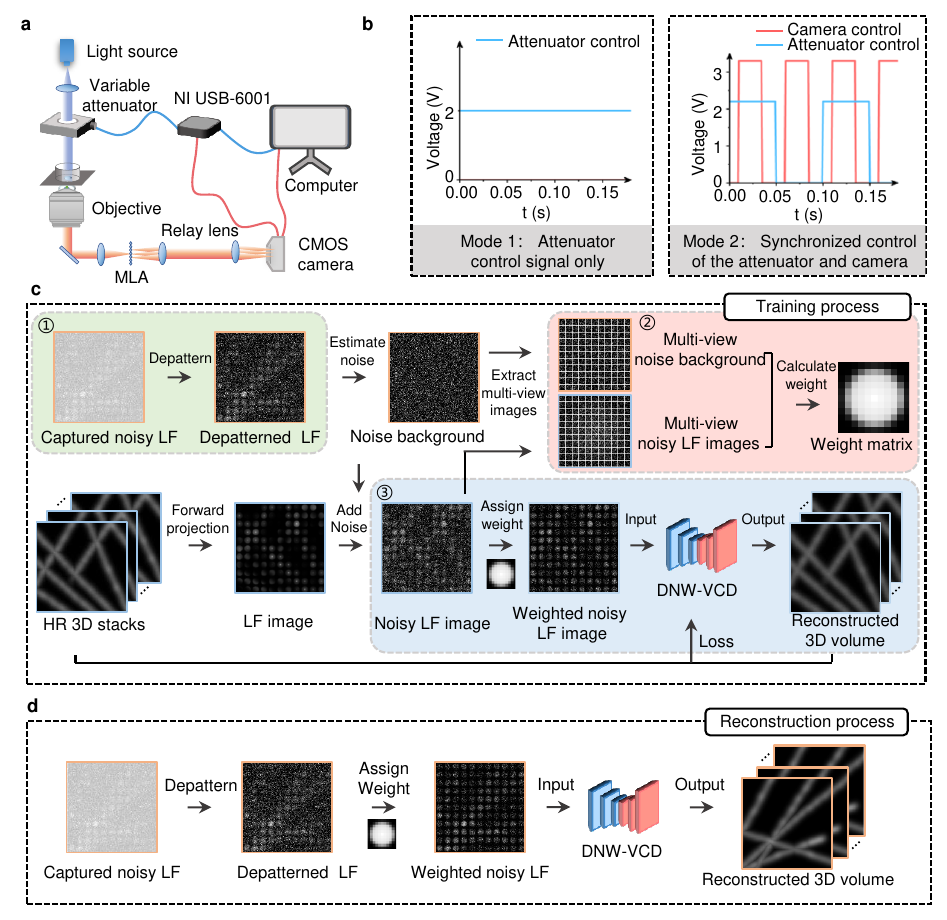}
\centering
\vspace{-0.2cm} %
\caption{ Hardware implementation and framework of DNW-VCD approach. (a) Optoelectronic setup of the light-field microscopy (LFM) system. A variable attenuator is added to the illumination path to modulate the excitation light intensity. The attenuator and the camera sensor are jointly controlled via a computer and a controller. (b) Control modes of the LFM imaging system. Mode 1 enables traditional light-field imaging with adjustable illumination intensity, while Mode 2 supports time-multiplexing high and low SNR light-field imaging. (c) DNW-VCD training workflow, including pattern noise estimation and removal (depattern), Gaussian noise estimation, weight matrix calculation using simulated pure-noise background and noisy light-field image pairs, training dataset generation, and network training. (d) DNW-VCD reconstruction workflow, involving the preprocessing step (depattern and weight assignment) and network-based inference.}
\label{fig1}
\vspace{-0.2cm} %
\end{figure}

The LFM system is constructed on an inverted fluorescence microscope (Zeiss, Axio Observer 7), as illustrated in Fig.~\ref{fig1}a. The microscope is equipped with a blue LED illumination source (Thorlabs, DC20) and a fluorescence illuminator (X-Cite 120Q) as the upright and inverted light sources, respectively. A PCO.panda 4.2 sCMOS camera (2048×2048 pixels and 6.5 \si{\micro\meter} pixel size) is employed for image acquisition. An MLA (RPC Photonics, MLA-S100-f21) is placed at the native image plane of the imaging system and relay lenses are used to conjugate the back focal plane of the MLA to the detection plane of the camera sensor. In experiments, a 20×/0.5 NA water immersion objective and a 40×/0.75 NA air immersion objective are used for imaging different samples.

An electronic-controlled variable attenuator (Thorlabs, LCC1620A) is added to the illumination path to modulate excitation light intensity (Fig.~\ref{fig1}a). By varying the voltage, the attenuator’s light transmittance could be adjusted. (1) Mode 1, traditional light-field imaging with adjustable illumination intensity: the attenuator voltage is set at fixed levels, producing stable excitation light. This allowed the acquisition of light-field images with varying noise levels by applying different voltages. (2) Mode 2, time-multiplexing high and low SNR light-field imaging: using square wave control signals, the attenuator and camera are synchronized via a USB NI-6001 interface. High SNR images are captured when the attenuator control signal is low, while low SNR images are acquired at high signal levels. Each cycle produces a time-multiplexed pair of high and low SNR light-field images, captured nearly simultaneously. These image pairs are used to validate the accuracy of the DNW-VCD approach in dynamic imaging scenarios.

By incorporating a two-step noise model and a viewpoint-dependent weight matrix into the VCD network architecture, \cite{RN17} a novel deep learning-based framework for real-time, artifact-reduced LFM reconstruction, our DNW-VCD approach achieves effective denoised reconstruction in LFM, even under extremely low-light conditions.

\subsubsection{Noise model of DNW-VCD}

In this work, the imaging noise is modeled as a fixed pattern noise from image sensor and a residual Gaussian noise.\cite{RN32} For a given pixel $i$ in a light-field image, its intensity value $D_{i}$ can be expressed as:

\begin{equation}
    D_i=\alpha_i\mathrm{P}(\lambda_i)+N(0,\sigma_R{}^2)+\beta_i,
    \label{}
\end{equation}
where $\lambda_{i}$ represents the photon count during exposure and $\mathrm{P}(\lambda_{i})$ is the corresponding imaging intensity of the pixel, following a Poisson distribution. $N(0,{\sigma_{R}}^{2})$ denotes the readout noise with a mean of 0 and a variance of $\sigma_R^2$. The fixed pattern noise, originated from the ACsN method,\cite{RN32} is parameterized by $\alpha_{i}$ (multiplicative) and $\beta_{i}$ (additive), which require a one-time calibration. For fluorescence microscopy scenarios, where $\lambda_i>3$, the Poisson distribution of photon shot noise can be approximated by a Gaussian distribution.\cite{RN32} Thus, the total noise is addressed in two steps: (1) fixed pattern noise removal and (2) modeling residual noise as the sum of two independent Gaussian random variables. The total noise variance becomes: ${\sigma_N}^2={\sigma_R}^2+{\sigma_G}^2$, where ${\sigma_R}^2$ and ${\sigma_G}^2$ represent the variances of the readout noise and photon shot noise, respectively. Our approach estimates ${\sigma_N}^2$ by leveraging the structural characteristics of light-field imaging. A detailed derivation of the noise model and the parameter calibration is provided in Supplementary. The forward imaging process of LFM is provided in Supplementary.

\subsubsection{Training process of DNW-VCD}

The DNW-VCD training process, as depicted in Fig.~\ref{fig1}c, includes several core steps. Details are also provided in Supplementary.

First, the fixed pattern noise of real-captured light-field images is computed and removed, as shown in Step \ding{172} of Fig.~\ref{fig1}c, referred to as depattern process in this work. After the depattern step, the remaining noise is considered as Gaussian noise. The variance ${\sigma_N}^2$ of this noise is estimated by only preserving and gathering pixels at the edges of the MLA, as detailed in Supplementary.

Second, as outlined in Step \ding{173} of Fig.~\ref{fig1}c, the weight matrix of different views in LFM is calculated. Noise-only background images are generated by adding Gaussian noise with the estimated parameters. Simulated noisy light-field images are created by forward-projecting the high-resolution ground-truth 3D stack and adding the same Gaussian noise. Multi-view images are then extracted from both the above pure-noise background and noisy light-field images to compute the weight matrix. For each view $j$, the weight $W_{j}$ in the weight matrix is calculated as:

\begin{equation}
W_{j} = \frac{E \cdot NLF_{j} - E \cdot N_{j}}{E \cdot NLF_{j}},
\label{eq:wj}
\end{equation}
where $E \cdot NLF_{j}$ and $E \cdot N_{j}$ denote the energy distribution of the noisy light-field image and the pure-noise background image, respectively.

Finally, as shown in Step \ding{174} of Fig.~\ref{fig1}c, the weighted noisy light-field images and high-SNR, high-resolution 3D stacks are input as training image pairs for the network. The weighted noisy light-field images are obtained by assigning the weight matrix into the simulated noisy light-field images. During training, the network minimizes the difference between the reconstructed 3D image and the high-resolution 3D stacks. The loss function, incorporating Maximum Intensity Projection (MIP) constraints\cite{RN20} is defined as:

\begin{equation}
\begin{split}
    {\mathcal{L}}^{(l)} = &\frac{1}{NV^{(l)}}\sum_{k=0}^{NV^{(l)}-1}\left(\hat{s}_{k}^{(l)}-s_{k}^{(l)}\right)^{2} \\
    &+ \frac{\gamma_{1}}{NXZ^{(l)}}\sum_{k=0}^{NXZ^{(l)}-1}\left(\mathcal{P}_{xz}\left\{\hat{s}_{k}^{(l)}\right\}-\mathcal{P}_{xz}\left\{s_{k}^{(l)}\right\}\right)^{2} \\
    &+ \frac{\gamma_{2}}{NYZ^{(l)}}\sum_{k=0}^{NYZ^{(l)}-1}\left(\mathcal{P}_{yz}\left\{\hat{s}_{k}^{(l)}\right\}-\mathcal{P}_{yz}\left\{s_{k}^{(l)}\right\}\right)^{2} \\
    &+ \frac{\gamma_{3}}{NXY^{(l)}}\sum_{k=0}^{NXY^{(l)}-1}\left(\mathcal{P}_{xy}\left\{\hat{s}_{k}^{(l)}\right\}-\mathcal{P}_{xy}\left\{s_{k}^{(l)}\right\}\right)^{2},
\end{split}
\end{equation}
where $s_k^{(l)}$ and ${\hat{s}}_k^{(l)}$ represent the $k$-th voxel of the ground truth and reconstructed distributions in the $l$-th training pair, respectively. $\mathcal{P}_{xz}\left\{\bullet\right\}$, $\mathcal{P}_{yz}\left\{\bullet\right\}$, and $\mathcal{P}_{xy}\left\{\bullet\right\}$ denote MIPs of the 3D object in the x-z, y-z, and x-y directions. ${NV}^{\left(l\right)}$ is the total voxel count of the 3D volume, while ${NXZ}^{\left(l\right)}$, ${NYZ}^{(l)}$, and ${NXY}^{(l)}$ represent the total pixel numbers of $\mathcal{P}_{xz}\left\{\bullet\right\}$, $\mathcal{P}_{yz}\left\{\bullet\right\}$, and $\mathcal{P}_{xy}\left\{\bullet\right\}$, with $\gamma_1$,$\ \gamma_2$,$\gamma_3$ as the corresponding weights.

\subsubsection{Reconstruction process of DNW-VCD}

The DNW-VCD reconstruction process, as shown in Fig.~\ref{fig1}d, involves removing the fixed pattern noise from real-captured light-field images (depattern process), assigning the calculated weight matrix to the depatterned light-field images, and feeding the weighted noisy light-field images into the trained DNW-VCD network for 3D volume reconstruction.

\section{Results}
In the experiments, the denoising capability of the proposed method is evaluated by comparing the reconstruction performance of four strategies: VCD, VCD-Noise, VCD-Depattern, and our proposed DNW-VCD. The VCD strategy utilizes the original VCD network to reconstruct the real-captured light-field images. The VCD-Noise strategy noise employs the VCD network trained with noisy light-field images to reconstruct the real-captured light-field images. The VCD-Depattern strategy applies the original VCD network to reconstruct the depatterned real-captured light-field images. DNW-VCD uses the VCD network trained with weighted noisy light-field images to reconstruct the depatterned real-captured light-field images. The VCD and VCD-Noise strategies are derived from the original VCD network\cite{RN17} and serve as the baseline comparisons. The VCD-Depattern strategy, proposed in this study, is employed for ablation experiments to verify the effectiveness of Gaussian noise estimation of residual noises and the weight matrix of multi-view images. Details are also provided in Appendix.

\begin{figure}[htbp]
\includegraphics[scale=0.75]{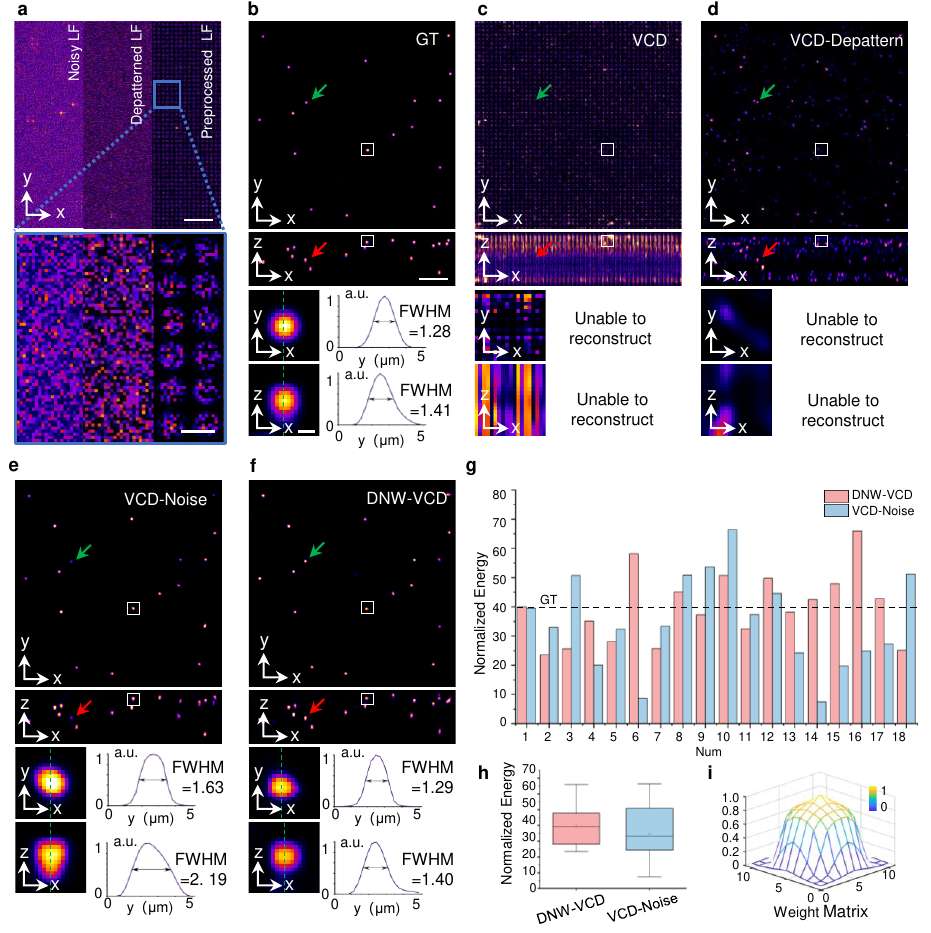}
\centering
\vspace{-0.2cm} %
\caption{ Performance and resolution evaluation of DNW-VCD by imaging simulated fluorescent beads. (a) Simulated noisy light-field images of fluorescent beads (1.28 \si{\micro\meter} diameter) using an air objective (40×/0.75 NA), along with light-field image after fixed pattern noise removal, preprocessed light-field image after weight application, and corresponding magnified details. (b) Maximum Intensity Projection (MIP) images of the ground truth. (c-f) Reconstructed results and magnified views using VCD, VCD-Depattern, VCD-Noise, and DNW-VCD strategies, illustrating lateral and axial MIPs, as well as the full width at half maximum (FWHM). a.u., arbitrary unit. (g) Energy statistics (normalized sum) of individual fluorescent beads (in e, f) reconstructed by DNW-VCD and VCD-Noise. The dashed line indicates the average ground truth energy. (h) Box plots showing the energy distribution of individual beads (in e, f) reconstructed by DNW-VCD and VCD-Noise. (i) Weight matrix utilized in this experiment. Scale bar, 20 \si{\micro\meter} (full FOV images in a), 3 $\si{\micro\meter}^{3}$ (magnified images in a), and 20 \si{\micro\meter} (full FOV images in b-f), 1 \si{\micro\meter} (magnified images in b-f).}
\label{fig2}
\vspace{-0.2cm} %
\end{figure}

\subsection{Characterization and evaluation of DNW-VCD}
To assess the reconstruction accuracy of DNW-VCD and evaluate the resolved axial and lateral resolutions, simulated fluorescent bead data are synthesized. Fluorescent beads with a diameter of 1.28 \si{\micro\meter} is modeled using a Gaussian filter and parameters corresponding to a 40× / 0.75 NA air objective. A simulated 3D distribution of beads is generated, followed by emulating the real-world imaging process of the light-field system and incorporating the noise distribution of camera acquisition. This resulting noisy light-field images have an approximate SNR of 2.0 dB.

The raw light-field images, depatterned light-field images, and DNW-VCD preprocessed light-field images are presented in Fig.~\ref{fig2}a. As shown in Fig.~\ref{fig2}b-f, the original VCD strategy fails to effectively reconstruct noisy light-field images, while VCD-Depattern strategy introduces significant artifacts. Although the VCD-Noise strategy accurately localizes the beads, as indicated by the green and red arrows in Fig.~\ref{fig2}e, it produces an uneven energy distribution in the reconstructed fluorescent beads. In contrast, the proposed DNW-VCD method achieves more stable and uniform denoising reconstruction, even under low SNR conditions. In the x-y and x-z MIP images shown in Fig.~\ref{fig2}b-f, the DNW-VCD method demonstrates superior lateral and axial resolution, with a particularly notable enhancement in axial resolution, achieving the highest imaging performance among the evaluated methods. To further evaluate the effectiveness of DNW-VCD compared to VCD-Noise, a statistical analysis is conducted on the normalized energy distribution of 18 reconstructed beads using both methods (see Fig.~\ref{fig2}g-h). The analysis reveals that DNW-VCD produces more uniform and stable reconstructions, highlighting its superior performance in low-SNR scenarios. Fig.~\ref{fig2}i illustrates the weight matrix used in this experiment.

Additionally, the reconstruction performance of DNW-VCD and VCD-Noise for fluorescent beads across various SNR levels is compared in Supplementary. The comparisons further demonstrate that DNW-VCD consistently outperforms VCD-Noise under different SNR conditions, highlighting its robustness. Ablation experiments are also conducted to examine the role of the weight matrix in the DNW-VCD method (see Supplementary).

\begin{figure}[htbp]
\includegraphics[scale=0.75]{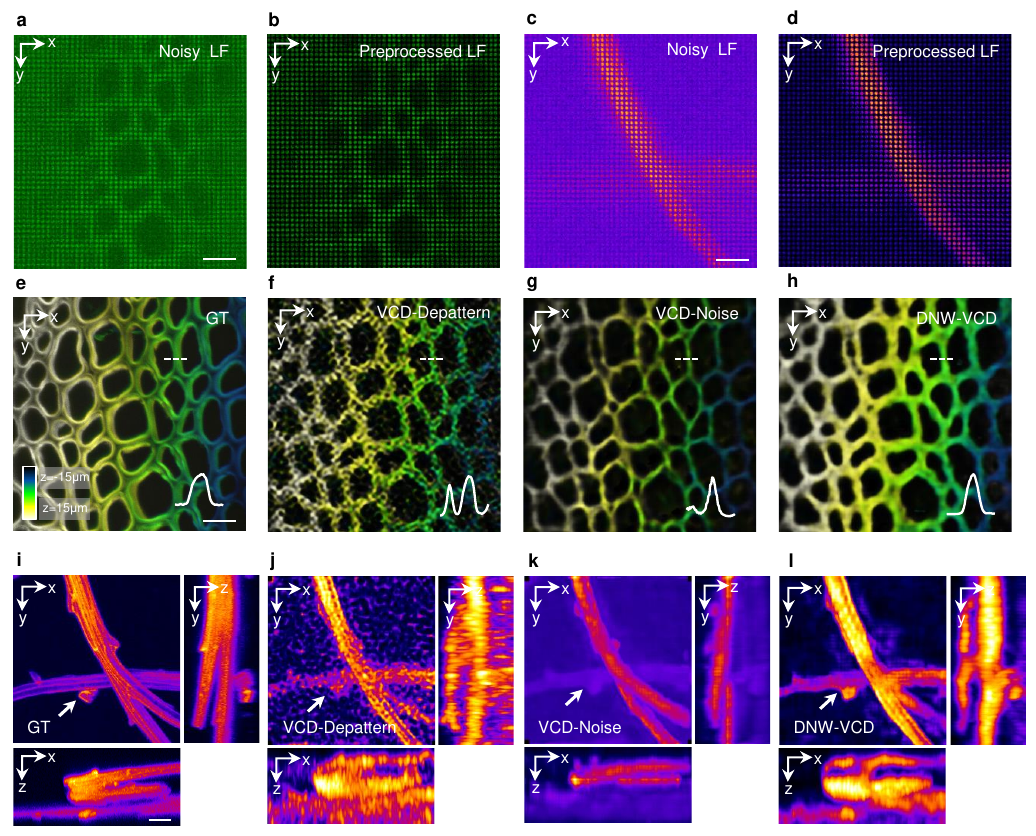}
\centering
\vspace{-0.2cm} %
\caption{ Performance of DNW-VCD by imaging biological samples. (a-b) Noisy light-field image of a cotton stem cross-section captured using Mode 1 of the LFM system with an air objective (40×/0.75 NA), and its corresponding preprocessed light-field image. (c-d) Noisy light-field image of a dandelion villi slice captured under the same condition, and its preprocessed light-field image. (e) Depth-color-coded lateral MIP image of the cotton stem cross-section obtained via confocal microscopy (Andor Technology, BC43, 20×/0.8 NA). (f-h) Depth-color-coded lateral MIP images of the noisy cotton stem cross-section reconstructed using VCD-Depattern, VCD-Noise, and DNW-VCD. Line profiles of the dash-line areas are displayed in the bottom-right corner of each panel. (i) Ground truth MIP images of the dandelion villi slice acquired with the same confocal microscopy. (j-l) Lateral and axial MIPs of the dandelion villi slice reconstructed using VCD-Depattern, VCD-Noise, and DNW-VCD. Scale bar, 20 \si{\micro\meter} (a-l).}
\label{fig3}
\vspace{-0.2cm} %
\end{figure}

\subsection{DNW-VCD achieves high-fidelity 3D reconstruction of low-SNR light-field data}

To validate the denoising capabilities of DNW-VCD for biological samples under real-world conditions, cotton stem cross-sections and dandelion villi slices are imaged using a 40×/0.75 NA air objective in Mode 1 of the system (see Methods). Light-field images are captured with low illumination intensity. The raw noisy light-field images are displayed in Fig.~\ref{fig3}a and \ref{fig3}c, while the DNW-VCD preprocessed images are presented in Fig.~\ref{fig3}b and \ref{fig3}d.

For the cotton stem cross-section (Fig.~\ref{fig3}e-h), the 3D reconstruction obtained using the VCD-Depattern strategy exhibits erosion-like artifacts in the sample structure caused by noise. The VCD-Noise strategy successfully reconstructs the general structure of the sample, but compared to the ground truth obtained by confocal microscopy (Andor Technology, BC43, 20× / 0.8 NA), the reconstructed result is too slim to accurately represent the true structure. In contrast, DNW-VCD delivers higher reconstruction quality, producing a 3D structure that is more similar to the ground truth. This improvement is evident in the line profiles of the dash-line areas, plotted in the bottom-right corner of each result.

Similar issues are observed with the dandelion villi slice (Fig.~\ref{fig3}i-l). The VCD-Depattern strategy results in a point-like, blocky mode in the 3D reconstruction, while the VCD-Noise strategy suffers from significant background interference and uneven energy distribution. In the region highlighted by the white arrows in Fig.~\ref{fig3}k, the VCD-Noise strategy fails to provide sufficient energy reconstruction, which causes the region to be overwhelmed by the background. In comparison, DNW-VCD significantly enhances energy uniformity, with both the lateral and axial reconstruction MIPs closely matching the ground truth. This experiment verifies the capability of DNW-VCD to achieve high-fidelity light-field reconstruction of biological samples under low-SNR conditions.

\subsection{DNW-VCD enables high-precision dynamic 3D imaging under low-light conditions}
To validate the denoising ability of DNW-VCD in dynamic 3D imaging under low-light scenes, we perform experiments involving the 3D tracking of fluorescence beads and the dynamic imaging of a zebrafish heart.

\subsubsection{Dynamic 3D tracking of fluorescence beads}
\begin{figure}[htbp]
\includegraphics[scale=0.82]{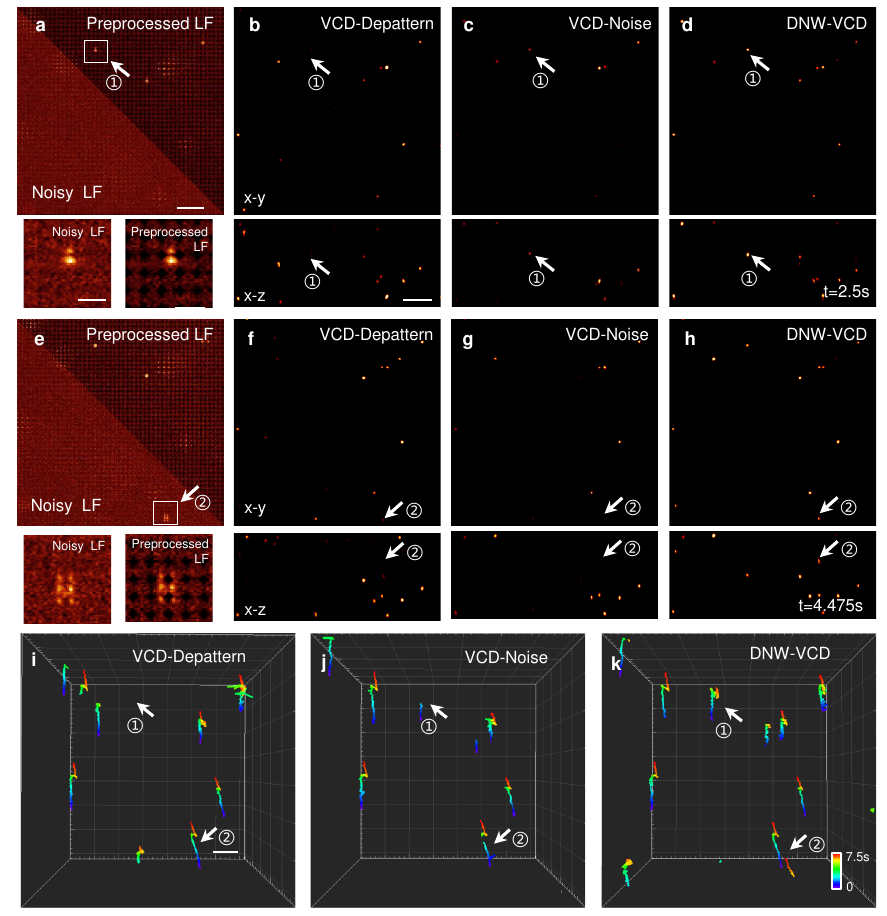}
\centering
\vspace{-0.2cm} %
\caption{ Dynamic 3D reconstruction capability of DNW-VCD for imaging floating fluorescent beads. (a) Noisy light-field image of fluorescent beads (diameter 2 \si{\micro\meter}) at t=2.500 s, captured in Mode 1 with a water immersion objective (20×/0.5 NA), along with the preprocessed image. (b-d) Lateral and axial MIPs of fluorescent beads at t=2.500 s, reconstructed using VCD-Depattern, VCD-Noise, and DNW-VCD. (e) Noisy light-field image of fluorescent beads at t=4.475 s, captured in Mode 1, and the preprocessed image. (f-h) Lateral and axial MIPs of fluorescent beads at t=4.475 s, reconstructed using VCD-Depattern, VCD-Noise, and DNW-VCD. (i-k) MIPs of fluorescent bead trajectories tracked by Imaris software, color-coded by time. Scale bar, 40 \si{\micro\meter} (full FOV image in a, e; b-d, f-k), 8 \si{\micro\meter} (magnified image in a, e).}
\label{fig4}
\vspace{-0.2cm} %
\end{figure}

Fluorescent beads with the diameter of 2 \si{\micro\meter} (Bitoyscience, ABTG-0200) are diluted and placed in the bottom of a glass dish (NEST, 15 mm diameter, 0.17 mm bottom thickness) to facilitate free movement in an aqueous medium. To minimize aberrations during imaging, the liquid droplet is flattened in the dish as much as possible and a water immersion objective (20× / 0.5 NA) is used for imaging. The space between the objective and the glass dish is filled with water to ensure proper immersion. Imaging is performed in Mode 1, and excitation light intensity is adjusted to ensure low-SNR conditions. A continuous 7.5-second video is recorded at a camera acquisition rate of 40 Hz. Light-field images at t = 2.500 s and t = 4.475 s are shown in Fig.~\ref{fig4}a and Fig.~\ref{fig4}e, respectively.

The 3D reconstruction of the captured light-field image series is performed using VCD-Depattern, VCD-Noise, and DNW-VCD strategies. As indicated by white arrows in Fig.~\ref{fig4}a-h, the VCD-Depattern strategy fails to reconstruct beads labeled \ding{172} and \ding{173} at both t = 2.500 s and t = 4.475 s. The VCD-Noise strategy is able to reconstruct bead \ding{172} but fails to resolve bead \ding{173}, and the energy distribution in reconstructed bead \ding{172} remains uneven. In contrast, DNW-VCD accurately resolves both beads, achieving superior reconstruction quality at both time stamps.

Furthermore, as shown in Fig.~\ref{fig4}i-k, the spatial trajectories of the fluorescent beads are tracked using the commercial Imaris software and color-coded images representing their motion over time are generated. The DNW-VCD strategy exhibits the best temporal continuity and 3D reconstruction quality among the tested strategies, as indicated by the color-coded images highlighted with white arrows. Additionally, the full width at half maximum (FWHM) of the reconstructed fluorescence beads is calculated (see Supplementary). We demonstrate that DNW-VCD yields the most uniform resolution and achieves the best lateral and axial resolution among the tested methods.

\subsubsection{Dynamic 3D imaging of zebrafish heart}
\begin{figure}[htbp]
\includegraphics[scale=0.82]{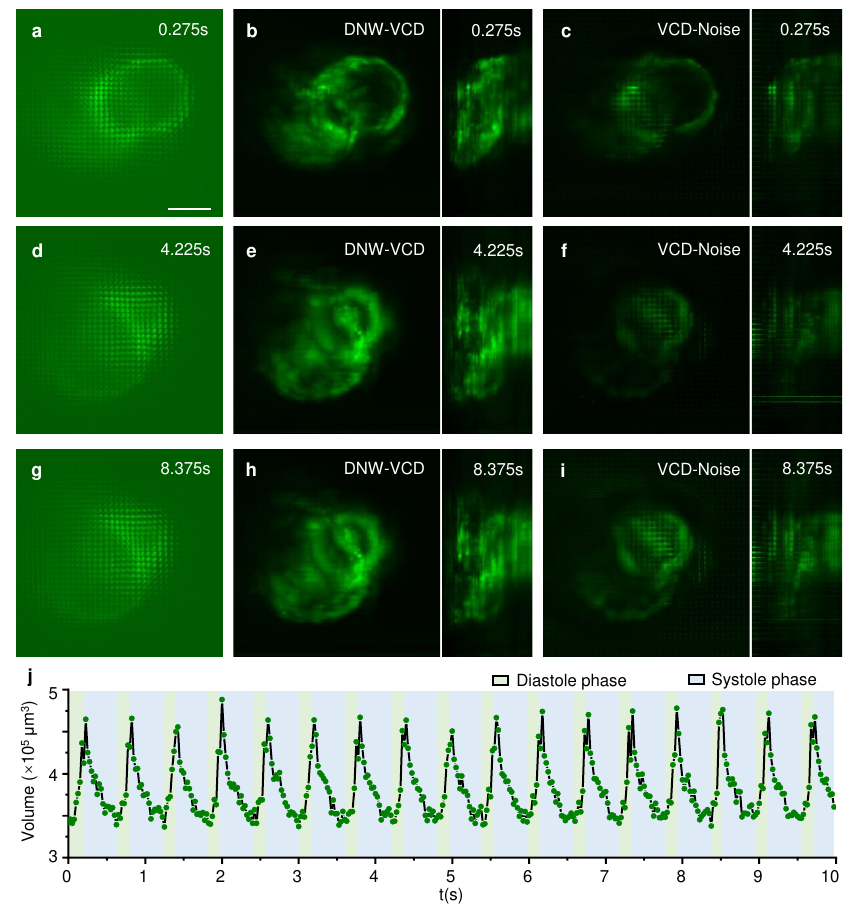}
\centering
\vspace{-0.2cm} %
\caption{ Dynamic 3D reconstruction capability of DNW-VCD for imaging living zebrafish heart. (a-c) Noisy light-field image of the zebrafish heart at t=0.275 s, imaged by a water immersion objective (20×/0.5 NA), along with its reconstruction results using DNW-VCD and VCD-Noise networks. (d-f) Noisy light-field image at t=4.225 s and its corresponding reconstruction results. (g-i) Noisy light-field image at t=8.375 s and its reconstruction results. (j) Volume variation curve of the zebrafish beating heart reconstructed with DNW-VCD over a 10-second duration capturing multiple cardiac cycles. Scale bar, 40 \si{\micro\meter}.}
\label{fig5}
\vspace{-0.2cm} %
\end{figure}

3D dynamic imaging of dense zebrafish heartbeats is performed using a water immersion objective (20×/0.5 NA). Living zebrafish larvae is placed at the bottom of a glass dish, anesthetized, and immobilized using 1\% low-melting-point agarose. Light-field sequences of the live zebrafish heart are captured under low-light conditions, covering 10-second duration at a frame rate of 40 Hz, with a volume of 200 × 200 × 85 $\si{\micro\meter}^{3}$. A publicly available dataset\cite{RN34} is utilized as the training set for both the DNW-VCD and VCD-Noise networks.

Captured light-field images are presented at three time points: t=0.275 s, t=4.225 s, and t=8.375 s in Fig.~\ref{fig5}a, \ref{fig5}d, and \ref{fig5}g, respectively. The denoised reconstruction obtained using DNW-VCD are shown in Fig.~\ref{fig5}b, \ref{fig5}e, and \ref{fig5}h, while the corresponding reconstructions using VCD-Noise are shown in Fig.~\ref{fig5}c, \ref{fig5}f, and \ref{fig5}i. The results indicate that DNW-VCD achieves higher denoising performance, effectively reducing artifacts and enhancing reconstruction quality. Additionally, the volume variation of the zebrafish heart over the 10-second period, encompassing multiple cardiac cycles, is plotted in Fig.~\ref{fig5}h. The average cardiac cycle duration is about 0.55 seconds, with diastole lasting 0.15 seconds and systole 0.40 seconds. A three-dimensional video depicting the entire zebrafish heart beating process is provided in Visualization 1. The above experiments collectively demonstrate that the proposed DNW-VCD, based on the intrinsic noise model of light fields, enables high-precision dynamic 3D imaging under low-light conditions.

\subsection{DNW-VCD enables high-fidelity real-time 3D imaging with 100-fold reduction in phototoxicity}

\begin{figure}[htbp]
\includegraphics[scale=0.75]{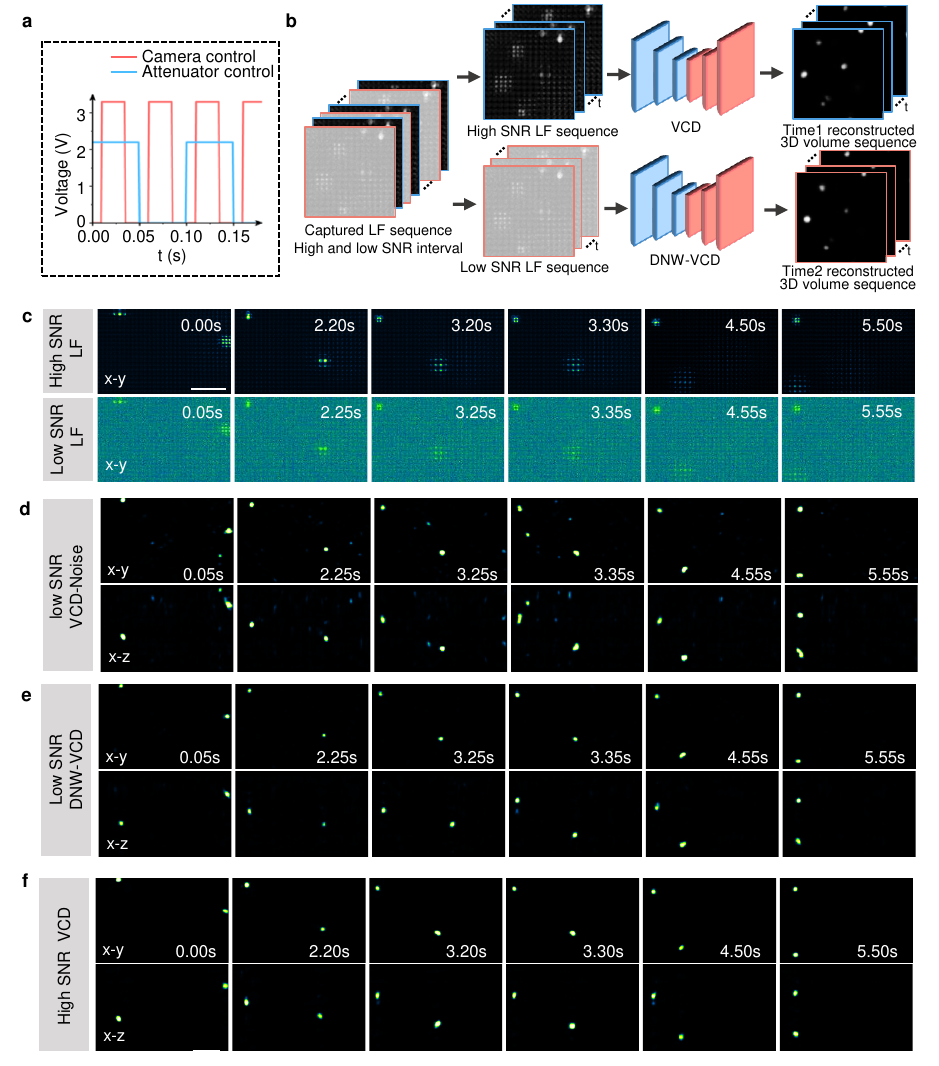}
\centering
\vspace{-0.2cm} %
\caption{Evaluation of DNW-VCD's Accuracy in dynamic 3D reconstruction of living algae under low phototoxicity. (a) Synchronized control signals given to the attenuator and camera in Mode 2. (b) Workflow for processing light-field images of live Dunaliella salina. (c) High- and low-SNR light-field sequences captured using a water immersion objective (20×/0.5 NA). (d-e) Lateral and axial MIPs of the low-SNR light-field sequence at different time points, reconstructed by DNW-VCD and VCD-Noise, respectively. (f) Lateral and axial MIPs of high-SNR light-field sequences at different time points, reconstructed by the original VCD. Scale bar, 40 \si{\micro\meter} (c-f).}
\label{fig6} 
\vspace{-0.2cm} %
\end{figure}

To demonstrate that DNW-VCD achieves high-fidelity real-time 3D imaging with significantly reduced phototoxicity, the dynamics of Dunaliella salina are imaged using a water-immersion objective (20× / 0.5 NA). The excitation light is adjusted via an attenuator with specific voltage settings (see Supplementary) to reduce phototoxicity, as shown in Fig.~\ref{fig6}a. The effective ratio of high-to-low light intensity is approximately 100-fold in our setup. Mode 2 of the system is employed to alternate between capturing high- and low-SNR frames of Dunaliella salina. The camera operates at a frequency of 20 Hz, while the attenuator control signal is set to 10 Hz. To accommodate the delay during the attenuator’s level transitions, a phase shift is introduced to the attenuator signal and its control frequency is reduced to 10 Hz. This ensures stable and uniform excitation light during image acquisition, resulting in a temporal sequence of alternating high- and low-SNR light-field images.

As shown in Fig.~\ref{fig6}b, the high-SNR light-field sequence is extracted and reconstructed using the VCD network as the ground truth, while the low SNR sequence is processed using the DNW-VCD network. Representative high- and low-SNR frame pairs of light-field image series are presented in Fig.~\ref{fig6}c. By comparing the reconstruction results of the two temporal image series, the denoising accuracy of DNW-VCD can be validated. The reconstruction of low-SNR light-field images using DNW-VCD (Fig.~\ref{fig6}e, \ref{fig6}f, and Visualization 2) closely aligns with the high-SNR reconstructions produced by the VCD network, effectively capturing Dunaliella salina dynamics while maintaining accurate lateral and axial scales. Conversely, reconstructions of low-SNR light-field images using the VCD-Noise strategy (Fig.~\ref{fig6}d), as a baseline for comparison, exhibits visible artifacts and noise. These findings highlight that DNW-VCD achieves superior performance in reconstructing low SNR light-field images, maintaining the fidelity and consistency with high-SNR light-field reconstructions while achieving about 100-fold reduction in phototoxicity.

To further evaluate DNW-VCD’s capability under low-light conditions with reduced phototoxicity, we utilize Mode 1 of the system to illuminate the algae samples and record the number of Dunaliella salina under both high- and low-light intensities during the same illumination period, enabling volumetric imaging at a scale of 250 × 250 × 101 $\si{\micro\meter}^{3}$ and a camera frame rate of 70 Hz. The experimental results are presented in Supplementary. The high-intensity, high-SNR images are reconstructed using VCD network, while the low-intensity, low-SNR images are processed using DNW-VCD. Starting with comparable algae counts in both groups, the low light group maintains higher algae count over time compared to the high light group. This result demonstrates that DNW-VCD, optimized for low-light, low-SNR 3D reconstruction, effectively minimizes sample damage and ensures superior bio-compatibility during imaging.

\section{Discussion and conclusion}

In this work, we propose the DNW-VCD network for LFM 3D reconstruction, integrating a two-step noise model and a weight matrix into the existing VCD framework. Our approach incorporates a noise estimation strategy based on the MLA mask characteristics and the intensity distribution of multi-view images, effectively modeling real-world noise distributions of LFM. We also develop the attenuator-induced imaging hardware for flexible illumination control, enabling simultaneous dual-SNR image acquisition. Through both simulations and experimental validation on biological samples, we demonstrate that DNW-VCD achieves uniform resolution, artifact suppression, and real-time 3D reconstruction even under extremely low-SNR conditions. DNW-VCD significantly enhances the performance of LFM for low-light 3D imaging, making it particularly well-suited for long-term 3D dynamic observations of biological samples while minimizing photobleaching and phototoxicity.

In LFM, sub-view images often exhibit low SNR under high-noise conditions, particularly at peripheral views, where even after denoising, some resolution loss persists. Our current approach attempts to address this by reducing the weights of peripheral views using the weight matrix, thereby minimizing reconstruction artifacts; however, this inevitably leads to some loss of information. To further optimize this, future efforts of this work could involve incorporating a sub-view enhancement network \cite{RN35} or developing advanced optical setups to improve peripheral view quality and overall reconstruction fidelity. Moreover, embedding noise estimation directly into the network structure could enable more convenient and efficient noise evaluation. Additional investigations may also explore integrating temporal constraints to enhance performance or employing self-supervised frameworks to reduce dependency on large-scale datasets during reconstruction.

\section*{Appendix A: Sample preparation and reconstruction dimensions}
\subsection*{Cotton stem cross-section and dandelion villi slice}
The two samples were purchased from AOXING Laboratory Equipment. After reconstruction, the voxel size was \(0.27 \times 0.27 \times 1 \, \si{\micro\meter}^3\). Due to the thickness of different samples, the reconstructed dandelion villi slice consisted of 41 axial slices, while the cotton stem cross-section included 31 axial slices.

\subsection*{Fluorescent beads}
The fluorescent beads, with a diameter of \(2 \, \si{\micro\meter}\) (Bitoyscience, ABTG-0200), were diluted and placed at the bottom of a glass dish (NEST, 15 mm in diameter and 0.17 mm bottom thickness), suspended in a water medium. To minimize aberrations during imaging, the thickness of the liquid droplet was flattened as much as possible. Imaging was conducted using a water immersion objective (\(20\times / 0.5 \, \text{NA}\)), and the space between the objective and the dish was also filled with water. The reconstructed fluorescent beads images consisted of 101 axial slices, with a voxel size of \(0.45 \times 0.45 \times 1 \, \si{\micro\meter}^3\).

\subsection*{Dunaliella salina}
Dunaliella salina were obtained from FENGYUN Algae Industry and placed the sample at the bottom of a glass dish for imaging. To minimize aberrations, a similar protocol was followed as for the fluorescent beads: a glass dish with a thickness of 0.17 mm was used for the algae sample, the droplet was flattened as much as possible, and a water immersion objective was employed for imaging. The space between the sample and the objective was also water-filled. The reconstructed algae images consisted of 101 axial slices, with a voxel size of \(0.45 \times 0.45 \times 1 \, \si{\micro\meter}^3\).

\subsection*{Zebrafish myocardium}
Transgenic zebrafish line Tg(cmlc2:eGFP) were used for dynamic zebrafish imaging. Zebrafish embryos were incubated at \(28.5^\circ\text{C}\) until they reached 2–3 days post fertilization (dpf). The zebrafish larvae were anesthetized using 0.16 mg/mL of 3-aminobenzoic acid ethyl ester methanesulfonate (\(\text{C}_9\text{H}_{11}\text{NO}_2 \cdot \text{CH}_4\text{O}_3\text{S}\), MESAB) and immobilized at the bottom of a glass dish with 1\% low-melting-point agarose solution. The morphology of the zebrafish was adjusted to maintain an upright position for cardiac imaging. Imaging was performed at room temperature. Ethical approval was not required for this experiment.

\section*{Appendix B: Data processing and simulation data generation}
\subsection*{Data processing}
In this work, we processed the data using MATLAB r2022b on a 64-bit machine equipped with an Intel Core i9-9900X CPU @ 3.50 GHz and 128 GB of memory. The training of the VCD, VCD-Noise, VCD-Depattern, and DNW-VCD networks was performed using TensorFlow 1.0 on an Nvidia GeForce RTX 2080 Ti GPU. The networks were trained on approximately 4,000 pairs of image blocks, with each pair consisting of a 3D volume of \(176 \times 176 \times (\text{number of axial slices})\) pixels as ground truth, and a corresponding \(176 \times 176\) pixel light-field images as the input. Training required 150 epochs and took approximately 7 hours on a single GPU. For the reconstruction of a single 3D volume, each network (VCD, VCD-Noise, VCD-Depattern, and DNW-VCD) took around 100 milliseconds to reconstruct a \(561 \times 561\) pixel LFM image.

\subsection*{Simulation data generation}
\subsubsection*{Simulated fluorescent bead}
We first simulated the generation of multiple binary beads with isotropic resolution, randomly distributed within a \(30 \, \si{\micro\meter}\) depth range in a 3D space. The data consisted of 31 axial slices, with a voxel size of \(0.27 \times 0.27 \times 1 \, \si{\micro\meter}^3\). Next, we applied a 3D Gaussian kernel to convolve the beads, controlling their FWHM as \(1.32 \, \si{\micro\meter}\). This approach produced a 3D stack of randomly distributed fluorescent beads, which we treated as the ground truth volume. One of these 3D stacks was selected as the sample for imaging, and we simulated the LFM imaging process by convolving the sample with the PSF, resulting in the raw light-field image free of noise contamination. To mimic the noise introduced during camera imaging, we added Gaussian noise and fixed pattern noise to the raw light-field image.

\subsubsection*{Simulated danio rerio cell membranes}
We selected an image from the publicly available dataset of Danio rerio cell membranes (cldnB:lyn-EGFP) as the sample for imaging \cite{RN36}. This dataset comprised 61 axial slices, with a voxel size of \(0.27 \times 0.27 \times 1 \, \si{\micro\meter}^3\). We convolved the dataset with the PSF to generate the light-field image and subsequently added noise to simulate the imaging process of the sample in LFM system.

\begin{backmatter}

\bmsection{Disclosures}
The authors declare no conflicts of interest.

\bmsection{Data Availability Statement}
The data that support the findings of this study are available from the corresponding author upon reasonable request.

\end{backmatter}

\bibliography{DNW-VCD_references}






\end{document}